\documentclass[article]{IEEEtran}
\IEEEoverridecommandlockouts
\usepackage{authblk}
\usepackage{datetime2}
\usepackage{graphicx}
\usepackage{amsmath}
\usepackage{amssymb}
\usepackage{mathrsfs}
\usepackage{stfloats}
\usepackage{dsfont}
\usepackage{setspace}
\usepackage{epstopdf}
\usepackage{cite}
\usepackage{balance}
\usepackage{xcolor}
\usepackage{booktabs}
\usepackage{pifont}
\usepackage[font=footnotesize]{caption}
\usepackage{subcaption}
\usepackage{bm}
\usepackage{bbm}
\usepackage[linesnumbered, ruled,vlined]{algorithm2e}
\usepackage{algorithmic}
\usepackage{sidecap}
\usepackage{hyperref}

\usepackage{fancyhdr}

\newtheorem{theorem}{Theorem}

\newtheorem{proposition}{Proposition}
\newtheorem{lemma}{Lemma}
\newtheorem{claim}{Claim}

\newtheorem{assumption}{Assumption}

\begin{document}
\title{
    Constrained MARL for Coexisting TN-NTN Resource Allocation: Scalability and Flexibility
}

\author{Cuong Le\authorrefmark{1}, Thang X. Vu\authorrefmark{1}, Stefano Andrenacci\authorrefmark{2}, and Symeon Chatzinotas\authorrefmark{1}

\thanks{\authorrefmark{1}Interdisciplinary Centre for Security, Reliability and Trust, University of Luxembourg, Luxembourg.}
\thanks{\authorrefmark{2} SES Satellites, Betzdorf, Luxembourg.}
\thanks{This work was supported in whole, or in part, by the Luxembourg National Research Fund, ref. C22/IS/17220888/RUTINE and BRIDGES/2023/IS/18441334/Pre5GNR.}
}

\markboth{Accepted to the 2026 IEEE International Conference on Communications}{}

\maketitle
\begin{abstract}
    This paper considers the joint TN-NTN constrained resource allocation, where terrestrial base stations and non-terrestrial base stations coexist in the spectrum. We focus on large-scale and practical scenarios characterized by large numbers of transmission channels and users, alongside highly dynamic user behaviors. As common learning solutions fail to address these challenges, we propose a decomposition solution based on the special properties of the cross-segment interference, and then tackle the original problem via solving subproblems in a sequential learning manner. Furthermore, to enhance the flexibility of the learned policies, we design a stochastic training environment that captures the key characteristics of real-world systems. Simulation results tested on the full 20MHz bandwidth with various numerologies show that our solution significantly improves scalability compared to existing solutions and remains robust in highly dynamic scenarios.
\end{abstract}

\section{Introduction}
\label{sec:introduction}
The 6\emph{th} Generation (6G) of the communications network involves multi-layer network architecture where both terrestrial networks (TNs) and non-terrestrial networks (NTNs) cooperatively serve the unprecedented demands for traffic and coverage. By combining terrestrial base stations with non-terrestrial platforms such as satellites and unmanned aerial vehicles (UAVs), TN-NTN systems promise enhanced coverage, connectivity, and flexibility in both urban and remote areas. However, the efficient allocation of resources in such hybrid systems is inherently complex, requiring the coordination of heterogeneous resources under dynamic and decentralized conditions. Traditional optimization methods, while theoretically sound, often fail to scale effectively in practical deployments due to their reliance on centralized computation. As 6G promises unparalleled capabilities driven by AI-powered automation, multiagent reinforcement learning (MARL) has emerged as a powerful paradigm for addressing the distributed and dynamic nature of TN-NTN systems.

In recent years, MARL has been increasingly applied for radio resource management, especially in channel assignment and power allocation \cite{tang2020deep, yang2022distributed, mei2024multi, li2022federated, liang2019spectrum, ji2023multi, naderializadeh2021resource, nasir2019multi,txvu2022}. Studies like \cite{tang2020deep, yang2022distributed, mei2024multi} demonstrated MARL for radio resource management in HetNets, where multiple agents are built to cooperatively perform channel assignment and power allocation. Similarly, MARL has been shown to be effective for managing spectrum sharing and transmit power in highly dynamic communication systems such as vehicular-to-vehicular (V2V) \cite{li2022federated, liang2019spectrum, ji2023multi}, cellular vehicular-to-everything (C-V2X)  \cite{hegde2023radio, guo2023radio}. Closer to our work are \cite{naderializadeh2021resource, nasir2019multi} which employ MARL with independent learning mechanism for resource management in cellular systems. A recent study in \cite{ji2024meta} proposes a framework which combines the ideas of meta-learning and federated learning to address the channel assignment and transmit power optimization problem, where PPO algorithm is used to train a pre-trained model at the base station to assign a user at most one subchannel. 
Despite showing promising results, studies have struggled to address the scalability challenge posed by the combinatorial nature of the channel assignment problem. Specifically, the exponential explosion of the action space renders MARL-based solutions fail to learn  effectively as the number of subchanels grows to the real-world configurations. This phenomenon, commonly known as the curse of dimensionality, refers to the degradation of algorithmic performance as the problem size grows. Existing studies either
exclude spectrum management to avoid this challenge~\cite{naderializadeh2021resource} or limit themselves to simulations on toy configurations with small numbers of transmission channels \cite{yang2022distributed}. 
Moreover, real-world systems are designed to be dynamic, allowing users to seamlessly join, move, and leave at any time, which is usually ignored in the literature \cite{yang2022distributed}. 
Despite the crucial importance of flexibility, there has been limited research explicitly addressing this issue. 

The above shortcomings motivate our research question in this study: \textit{Can we create a scalable and universal policy that generalizes to the dynamic scenarios of real-world TN-NTN systems?} In the attempt to address this question, \textit{i})~we consider the joint TN-NTN resource allocation problem, where terrestrial base stations (TBSs) and non-terrestrial base station (NTBSs) coexist in spectrum. Our focus is on practical scenarios characterized by large numbers of sub channels and users, alongside highly dynamic user behaviors. \textit{ii}) Based on special properties of the system, we propose a decomposition method to break the original problem into subproblems, each of which is easier to solve and, more importantly, the optimality of the sub-solutions is jointly preserved in the
original problem. \textit{iii}) We employ the Multi-Agent Proximal Policy Optimization (MAPPO) algorithm to solve the subproblems and use the technique of Lagrangian relaxation to handle the QoS constraints. \textit{iv}) In addition, we also design a stochastic training environment to enhance the flexibility of the learned policies. Experimental results demonstrate that our solution significantly enhances scalability compared to existing solutions and remains robust when tested in highly dynamic scenarios of real-world systems.

\section{System Model}
\label{sec:model}
We consider a downlink system with a set $\mathcal{N}$ of $N$ base stations (BSs), comprising both TBSs and NTBSs deployed either temporarily for rapid coverage extension or as part of an integrated network. Let $\mathcal{K}_n$ be the set of $K_n$ users associated with BS $n$ and $\mathcal{K} = \cup_{n\in\mathcal{N}} \mathcal{K}_n$ be the set of all users. BSs are assumed to share the same frequency band $B$, which is divided into $F$ subchannels, each with bandwidth of $W$. Let $h_{k,n,f}$ be the channel coefficient of subchannel $f$ between BS $n$ and user $k$. In the case where BS $n$ is a NTBS, it is likely that $h_{k,n,f}$ represents a line-of-sight channel. Otherwise, $h_{k,n,f}$ is assumed to follow a Rayleigh fading channel. In order to mitigate intra-BS interference, we employ OFDM transmission as follows. Let $x_{k,n,f}$ be a binary variable that captures the assignment of subchannel $f$ of BS $n$ to its associated user $k \in \mathcal{K}_n$, and let $p_{k,n, f}$ be the corresponding transmit power. We first have the orthogonal constraint: \begin{equation}
    {\sum}_{k\in\mathcal{K}_n}x_{k,n,f} \leq 1, \forall f\in\mathcal{F}_n, n\in\mathcal{N} \label{ctr:orthogonal}
\end{equation}
where $\mathcal{F}_n$ is the index set of all subchannels of BS $n$. By imposing the following constraint on the transmit power,
\begin{equation}
    p_{k,n,f} \leq x_{k,n,f}p_{\max}, \forall k\in\mathcal{K}_n, f\in\mathcal{F}_n, n\in\mathcal{N}\label{ctr:sub_channel_power}
\end{equation}
the achievable rate for user $k$ on the subchannel $f$ of BS $n$ can be written as
\begin{equation}
    R_{k, n, f} = W\log_2\Big(1 + \frac{|{h}_{k, n,f}|^2{p}_{k,n, f}}{g_{k,n,f} + WN_0}\Big) \label{eq:rate}
\end{equation}
where $g_{k,n,f} = \sum_{l\in\mathcal{N}\backslash\{n\},k'\in\mathcal{K}_l}|h_{k,l,f}|^2 p_{k',l,f}$ is the inter-BS interference power and $N_0$ is the noise density.
\subsection{Conventional optimization formulation}
The problem of maximizing achievable sum-rate can be formulated as
\begin{align}
    \max_{\textbf{x}, {\textbf{p}}}  \quad& {\sum}_{n\in\mathcal{N}}{\sum}_{k\in\mathcal{K}_n}{\sum}_{f\in \mathcal{F}_n}R_{k, n, f}\label{prob:max_sum_rate}\tag{\sf P1} \\
    \text{s.t.} \quad
    &{\sum}_{f\in \mathcal{F}_n}R_{k, n, f} \geq \eta_k, \forall k\in \mathcal{K}_n, n\in \mathcal{N}\label{ctr:qos}\\
    &{\sum}_{f\in\mathcal{F}_n} {\sum}_{k\in\mathcal{K}_n}p_{k,n, f} \leq p_{\max},\forall n\in\mathcal{N}\label{ctr:total_power}\\
    &\eqref{ctr:orthogonal},\eqref{ctr:sub_channel_power},\nonumber
\end{align}
where $\bold{x}$ and $\bold{p}$ are the shorthand notations of channel assignment and power allocation variables. 
Solving \eqref{prob:max_sum_rate} by centralized optimization \cite{FatemehTWC25} is impractical for real-world systems due to i) the lack of global CSI and ii) excessive signaling overhead and latency. Similarly, decentralized optimization suffers from limited inter TN-NTN coordination, leading to ineffective interference management and possible QoS violations. In practice, the 3GPP standardization only allows a BS to estimate the CSI to its connected users, making it is impossible to solve problem \eqref{prob:max_sum_rate} optimally.

\subsection{Standard Dec-POMDP Reformulation}
The most common approach to convert \eqref{prob:max_sum_rate} into an Decentralized partially observable Markov decision process (Dec-POMDP) is to consider each BS as an agent, which controls the corresponding set of subchannels \cite{tang2020deep, yang2022distributed, mei2024multi,naderializadeh2021resource, nasir2019multi}. Accordingly, the problem can be represented by a tuple $\langle\mathcal{N}, \mathcal{S}, \mathcal{A}, P, \mu, r, c, \mathcal{Z}, \gamma\rangle$, where $\mathcal{N}$ is the set of all agents, $\mathcal{S}$ is the state space, $\mathcal{A} = \prod_{n\in\mathcal{N}}\mathcal{A}_{n}$ is the joint action space, $\mathcal{A}_{n}$ is the action space of agent $n$ controlling BS $n$, $P:\mathcal{S}\times\mathcal{A}\times\mathcal{S}\rightarrow\mathbb{R}$ is the environment transition kernel, $\mu$ is the initial state distribution, $r:\mathcal{S}\times\mathcal{A}\rightarrow\mathbb{R}$ is the reward function shared by all agents, $c:\mathcal{S}\times\mathcal{A}\rightarrow\mathbb{R}^K$ is the cost function used for QoS constraint handling, $\mathcal{Z} = \prod_{n\in\mathcal{N}}\mathcal{Z}_{n}$ is the joint observation space, and $\gamma$ is the discount factor.

In time step $t$, let $a_{t,n}\in \mathcal{A}_{n}$ be the action of agent $n$ and $a_t\in \mathcal{A}$ be the joint action of all agents. Similarly, let $z_{t,n}\in\mathcal{Z}_{n}$ be the observation of agent $n$ and $s_t\in\mathcal{S}$ be the global state.
Let $H_{t,n} \in (\mathcal{Z}_n\times\mathcal{A}_n)^{t-1}\times \mathcal{Z}_n$ be the action-observation history of agent $n$ in the time step $t$ given by $H_{t,n} = (z_{0,n}, a_{0,n}, z_{2,n}, \ldots, a_{t-1,n}, z_{t,n})$.
Let $r(s_t, a_t)$ be the reward obtained by taking the joint action $a_t$ in state $s_t$, and $c_k(s_t, a_t)$ be the corresponding cost associated with user $k$.  
Let $\pi_{n}:(\mathcal{Z}_n\times\mathcal{A}_n)^{t-1}\times \mathcal{Z}_n\rightarrow \Delta_{|\mathcal{A}_n|}$ be the policy of agent $n$ at BS $n$, where $\Delta_{|\mathcal{A}_n|}$ is a probability simplex of dimension $|\mathcal{A}_n|$.
We are to find a joint policy $\pi$, where $\pi(a_t|s_t) = \prod_{n\in\mathcal{N}}\pi_n(a_{t,n}|s_t)$, to maximize the expectation of total discounted reward while satisfying the QoS, expressing via a cost parameter $d_k$:
\begin{align}
    \underset{\pi}{\text{Maximize}}\quad\quad & \mathbb{E}_{\mu, \pi}\left[{\sum}_{t=0}^{\infty}\gamma^tr(s_t, a_t)\right]\label{prob:cmdp}\tag{\sf RL1}\\
    \text{subject to}\quad\quad
    &c_{k}(s_t, a_t) \leq d_k, \forall k\in \mathcal{K}, t\in\mathbb{N}.
\end{align}
The above standard formulation works relatively well for small instances. However, for practical scenarios, e.g., hundreds subchannels and dozen of users per BS, solving \eqref{prob:cmdp} becomes challenging due to the combinatorial explosion of the action space. Therefore, conventional learning algorithms cannot be used without specific designs of the decision process. In the next subsection, we propose a Dec-POMDP formulation with decomposition that \emph{exploits the structural properties} of problem \eqref{prob:max_sum_rate} to break it down into simpler subproblems. 

\section{Proposed Framework}
\label{sec:solution}
\subsection{Proposed Formulation with Decomposition}
Our goal is to reformulate \eqref{prob:cmdp} into multiple subproblems, each of which is easier to solve, and more importantly, the performance loss caused by the decomposition is minimal. To achieve this, instead of training a single agent for each BS to manage all subchannels as in the conventional formulation, we employ multiple agents for each BS, with each agent responsible for only a subset of subchannels. Without loss of generality, assume that the subchannels of BS $n$ are divided into $M$ subsets denoted by ${\mathcal{F}_{n,1},\mathcal{F}_{n,2},\ldots, \mathcal{F}_{n,M}}$, where, $\forall m_1\neq m_2 \text{ and } n_1, n_2\in\mathcal{N}$,
\begin{equation}
    \cup_{m=1}^M\mathcal{F}_{n, m} = \mathcal{F}_n \text{ and }\mathcal{F}_{n_1, m_1}\cap \mathcal{F}_{n_2, m_2}=\emptyset.\label{eq:decomposition}
\end{equation}
For ease of presentation, let call each subset $\mathcal{F}_{n,m}$ a \textit{subset resource block} (SRB). To enable parameter sharing between agents, the above division is performed in such a way that all SRBs have the same cardinality. Our problem now is to train $M$ agents for each BS, where agent $m$ of BS $n$, denoted by $A_{n,m}$, is responsible for the SRB $\mathcal{F}_{n,m}$. 

\begin{figure}
    \centering
    \includegraphics[width=0.8\linewidth]{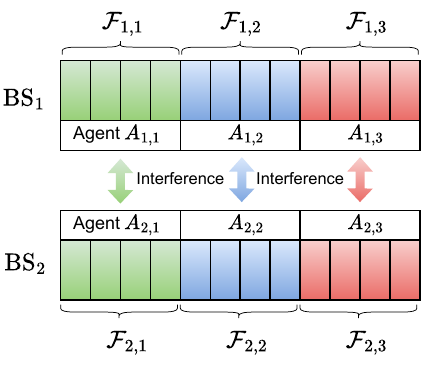}
    \caption{Illustration of the decomposed problem with BSs and three SRBs, with each SRB at a BS controlled by one agent.}
    \label{fig:decomposition}
\end{figure}

The rationale behind the above decomposition stems from the fact that although the number of subchannels can be large, interference occurs only between a few subchannels of different BSs that occupy the same frequency band. As a consequence, for any $m_1 \neq m_2$, there is no interference between subchannels in $\mathcal{F}_{n_1, m_1}$ and those in $\mathcal{F}_{n_2, m_2}, \forall n_1, n_2$. This enables us to group all agents in the system into $M$ groups, where interference management is only necessary between $N$ agents of each group.
In Fig. \ref{fig:decomposition}, we illustrate our decomposition with two BSs controlled by six agents.

Let $\mathcal{M}_n = \{A_{n,1}, A_{n,2},\ldots, A_{n,M}\}$ be the set of $M$ agents of BS $n$. Let $\mathcal{G}_m = \{A_{1,m}, A_{2,m}, \ldots, A_{N,m}\}$ be the set of $N$ agents of the agent group $m$. In order to maximize the sum data rate while meeting the QoS requirements, each agent $A_{n,m}$ only needs to cooperate with agents of the same group $\mathcal{G}_m$ to mitigate interference, while simultaneously cooperating with agents in $\mathcal{M}_n$ to guarantee the QoS at BS $n$. To achieve this interaction model, we let the agents of each group share a common reward function, while the agents of each BS \textit{make decisions sequentially}, allowing each agent to adjust its action by considering the actions of others (of the same BS) to satisfy the QoS constraints. 

Following the above idea, the decomposed problem is formulated as follows. To avoid notation burden, we will reuse some symbols from the previous subsection, adding an overbar when necessary. The problem is represented as a Markov game $\langle\mathcal{M}, \mathcal{S}, {\mathcal{A}}, P, \mu, \Bar{r}, c, {\mathcal{Z}}, \gamma\rangle$, where $\mathcal{M} = \bigcup_{n\in\mathcal{N}}\mathcal{M}_n$ is the set of all agents, ${\mathcal{A}} = \prod_{n\in\mathcal{N},m\in\mathcal{M}_n}\Bar{\mathcal{A}}_{n,m}$ is the joint action space with $\Bar{\mathcal{A}}_{n,m}$ denoting the action space of agent $A_{n,m}$, $\Bar{r}$ is now a set of $M$ reward functions where $\Bar{r}_m:\mathcal{S}\times(\prod_{m\in\mathcal{M}_n}\Bar{\mathcal{A}}_{n,m})\rightarrow\mathbb{R}$ is the reward function shared by $N$ agents of group $m$, ${\mathcal{Z}} = \prod_{n\in\mathcal{N},m\in\mathcal{M}_n}\Bar{\mathcal{Z}}_{n,m}$ is the joint observation space where $\Bar{\mathcal{Z}}_{n,m}$ is observation space of $A_{n,m}$.

\begin{figure}
    \centering
    \includegraphics[width=0.95\linewidth]{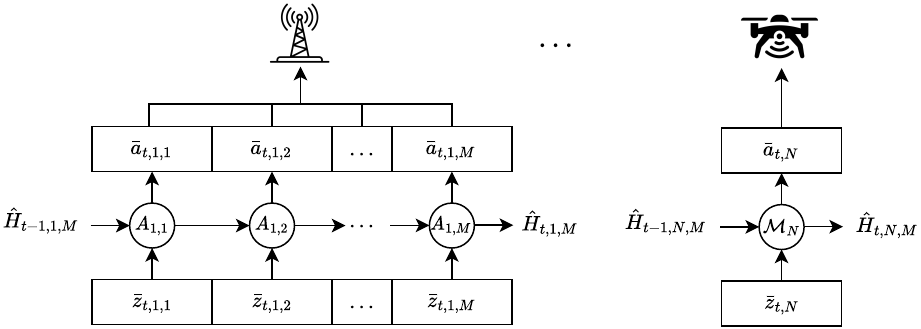}
    \caption{The computational model of the proposed solution, where the action of each BS in each time step is calculated sequentially by its $M$ agents.}
    \label{fig:computation_model}
\end{figure}

Without loss of generality, assume that, in time step $t$, the first $m-1$ agents of BS $n$ have made their decisions. Let $\Bar{z}_{t,n,m}\in\Bar{\mathcal{Z}}_{n,m}$ and $\Bar{a}_{t,n, m}\in \Bar{\mathcal{A}}_{n, m}$ be the observation and action of agent $A_{n,m}$ in time step $t$. Let define $\Bar{H}_{t, n, m} \in \mathcal{H}_{n,m}$, with $\mathcal{H}_{n,m} = \prod_{i=1}^{m-1}(\Bar{\mathcal{Z}}_{n,i}\times \Bar{\mathcal{A}}_{n, i})\times \Bar{\mathcal{Z}}_{n,m}$, as the action-observation step-history of BS $n$ in step $t$ up until the decision-making of agent $A_{n,m}$, $\Bar{H}_{t, n, m} = (\Bar{z}_{t,n,1}, \Bar{a}_{t,n,1}, \Bar{z}_{t,n,2}, \ldots, \Bar{a}_{t,n,m-1}, \Bar{z}_{t,n,m}).$
Let $\hat{{H}}_{t,n,m}\in  (\prod_{i=1}^{M}\Bar{\mathcal{Z}}_{n,i}\times \Bar{\mathcal{A}}_{n, i})^{t-1}\times  \mathcal{H}_{n,m}$ be the entire action-observation history of all agents of BS $n$ up until the decision-making of agent $A_{n,m}$ in time step $t$, $\hat{{H}}_{t,n,m} = (\Bar{H}_{0,n, M}, \Bar{a}_{0,n,M}, \ldots, \Bar{H}_{t-1,n, M},\Bar{a}_{t-1,n,M}, \Bar{H}_{t,n, m}).$
Let $\Bar{\pi}_{n,m}:(\prod_{i=1}^{M}\Bar{\mathcal{Z}}_{n,i}\times \Bar{\mathcal{A}}_{n, i})^{t-1}\times \mathcal{H}_{n,m}\rightarrow \Delta_{|\Bar{\mathcal{A}}_{n,m}|}$ be the policy of agent $A_{n,m}$, which maps $\hat{{H}}_{t,n,m}$ to a point of the probability simplex $\Delta_{|\Bar{\mathcal{A}}_{n,m}|}$. Let $\Bar{a}_{t,m} = (\Bar{a}_{t,1,m}, \Bar{a}_{t,2,m},\ldots, \Bar{a}_{t,N,m})$ be the joint action of agent group $\mathcal{G}_m$ and $\Bar{a}_{t}$ be the joint action of all agents. Let $\Bar{r}_m(s_t, \Bar{a}_{t,m})$ be the shared reward value obtained by all agents in $\mathcal{G}_m$ for taking the joint action $\Bar{a}_t$ in state $s_t$. In time step $t$, let $\Bar{a}_{t,n,1:m-1}$ be the joint action of the first $m-1$ agents of BS $n$ and $\Bar{a}_{t,1:m-1}$ be the joint action of the first $m-1$ agent groups. Let $\Bar{\pi}_{1:m-1}$ be the joint policy of the first $m-1$ agent groups. For each agent group $\mathcal{G}_m$, the problem is to find the joint policy $\Bar{\pi}_m$, where $\Bar{\pi}_m(\Bar{a}_{t,m}|s, \Bar{a}_{t,1:m-1}) =  \prod_{n\in\mathcal{N}}\Bar{\pi}_{n,m}(\Bar{a}_{t,n,m}|s, \Bar{a}_{t,1:m-1})$, to maximize the expectation of total discounted reward why satisfying the QoS constraints:
\begin{align}
    \underset{\Bar{\pi}_m}{\text{Max}}\quad &\mathbb{E}_{\mu, \Bar{\pi}_m}\left[{\sum}_{t=0}^{\infty}\gamma^t\Bar{r}_m(s_t, \Bar{a}_{t,m})|\Bar{\pi}_{1:m-1}\right]\label{prob:cmdp_decomposed}\tag{\sf RL2}\\
    \text{s.t.}\quad
    &c_{k}(s_t, \Bar{a}_{t}) \leq d_k, \forall k\in \mathcal{K}, t\in\mathbb{N}\label{ctr:qos_cmdp_decomp}.
\end{align}
Here, the conditional expectation in the objective of \eqref{prob:cmdp_decomposed} and conditional probability in the policies are due to the sequential decision-making process as discussed previously. From the solutions of subproblems \eqref{prob:cmdp_decomposed}, we have the solution of \eqref{prob:cmdp} given by $\Bar{\pi}=\prod_{m=1}^M\Bar{\pi}_m$. In Fig. \ref{fig:computation_model}, we illustrate the computational model of the proposed solution.

\subsection{Relation with the Original Formulation}
In this subsection, we establish the relation the subproblem \eqref{prob:cmdp_decomposed} and the original problem \eqref{prob:cmdp}. Since the QoS constraint in \eqref{ctr:qos_cmdp_decomp} is imposed on the joint action of all agent groups, if the subpolicy $\Bar{\pi}_m$ is feasible to \eqref{prob:cmdp_decomposed}, then the joint policy is feasible to \eqref{prob:cmdp}. For simplicity, the following analyses are derived assuming that \eqref{ctr:qos_cmdp_decomp} is satisfied. 
\begin{assumption}
    \label{assumption:reward_invariant}
    For any joint action $\Bar{a}_{t,m}$ of agent group  $\mathcal{G}_m$,
    $\Bar{r}_m(s_t, \Bar{a}_{t, m})$ is independent of policy $\Bar{\pi}_{m'}, \forall m'\neq m, t\in\mathbb{N}$.
\end{assumption}
\begin{assumption}\label{assumption:reward_decomposition}
      For any joint action $\Bar{a}_t = (\Bar{a}_{t,1}, \ldots, \Bar{a}_{t,M})$, $r(s_t, \Bar{a}_t) = \sum_{m=1}^M\Bar{r}_m(s_t, \Bar{a}_{t,m})$.
\end{assumption}
\noindent 
The first assumption posits that, while the actions of previous $m-1$ agent groups may influence the decisions and received reward of agent group $m$, they do not affect the agent group $m$'s underlying reward function.
The second assumption implies that the reward of a joint action in the original problem \eqref{prob:cmdp} can be additively decomposed into the rewards of constituent actions in subproblems. 
We will show later that the above assumptions can be satisfied by properly designing the reward functions and action space. From these assumptions, we have the following useful lemmas:
\begin{lemma}\label{lemma:obj_decomposition}
    Let $\Bar{J}_m(\Bar{\pi}_m|\Bar{\pi}_{1:m-1})$ be the objective value of \eqref{prob:cmdp_decomposed} at the policy $\Bar{\pi}_m$ w.r.t. $\Bar{\pi}_{1:m-1}$, and $J({\Bar{\pi}})$ be the objective value of \eqref{prob:cmdp} at the joint policy $\Bar{\pi}=\prod_{m=1}^M\Bar{\pi}_m$. Then $J(\Bar{\pi}) =\sum_{m=1}^M\Bar{J}_m(\Bar{\pi}_m|\Bar{\pi}_{1:m-1}).$
\end{lemma}
\begin{lemma}
\label{lemma:invariant_V^*_m}
    For each agent group $m$, let $\Bar{\pi}^*_m$ be the optimal policy w.r.t. a joint policy $\Bar{\pi}_{1:m-1}$ of the first $m-1$ agent groups, and let $\Bar{J}^*_m$ be the corresponding optimal value of \eqref{prob:cmdp_decomposed} under $\Bar{\pi}^*_m$. Then, $\Bar{J}^*_m$ is unique for all $\Bar{\pi}_{1:m-1}$.
\end{lemma}
\noindent While lemma \ref{lemma:obj_decomposition} is derived based on assumption \ref{assumption:reward_decomposition} and the linearity of expectations, lemma \ref{lemma:invariant_V^*_m} is obtained by transforming \eqref{prob:cmdp_decomposed} into a linear program and then invoking assumption \ref{assumption:reward_invariant}. From these lemmas, we obtain the main result as follows:
\begin{theorem}[Optimality preservation]
    Assuming that constraint \eqref{ctr:qos_cmdp_decomp} is satisfied, a policy $\Bar{\pi}^*_m$ is an optimal solution to \eqref{prob:cmdp_decomposed} for $m=1,2,\ldots,M$, iff the joint policy $\Bar{\pi}^*$, where $\Bar{\pi}^*(\Bar{a}_t|s_t) = \prod_{m=1}^M\Bar{\pi}^*_m(\Bar{a}_{t,m}|s_t,\Bar{a}_{t,1:m-1})$, is optimal solution to \eqref{prob:cmdp}.
\end{theorem}

The proof of Theorem~1 is omitted due to space limitation.

One important observation is that, compared to the original problem \eqref{prob:cmdp}, the joint action space for channel assignment in \eqref{prob:cmdp_decomposed} is reduced by a factor of $K^\frac{N(M-1)F}{M}$.
This can be verified by noting that each subproblem \eqref{prob:cmdp_decomposed} has its own reward function. In summary, we have decomposed the original problem \eqref{prob:cmdp} into $M$ subproblems, each of which has a smaller action space and is thus easier to solve. More importantly, if we can find optimal policies of all subproblems, it is guaranteed that the joint policy is optimal for the original problem. 

\subsection{Details of the Proposed Algorithm}
The analysis derived in the preceding subsection is optimistic in the sense that it does not account for QoS constraints. Nevertheless, this theoretical result offers valuable insights for designing the decision process to minimize performance loss due to decomposition. Specifically, to benefit from this result, the design must satisfy our two assumptions.

\subsubsection{Reward function} Since the evolution of channel state information is independent from past actions of the agents, maximizing the expected total data rate of a long-term horizon is equivalent to maximizing this of every time step as in \eqref{prob:max_sum_rate}. This implies that to achieve the objective of \eqref{prob:max_sum_rate}, the total data rate can be used as the reward function of \eqref{prob:cmdp} as follows
\begin{equation}
    r(s_t, a_t) = \frac{1}{K}{\sum}_{n\in\mathcal{N}}{\sum}_{f\in\mathcal{F}_{n}}{\sum}_{k\in\mathcal{K}_n}R_{k,n,f}(s_t, a_t).
\end{equation}
where $R_{k,n,f}(s_t, a_t)$ represents the data rate of user $k$ on the subchannel $f$ of BS $n$ resulting from taking the joint action $a_t$ in state $s_t$. For \eqref{prob:cmdp_decomposed}, we define the reward function as the total data rate on the corresponding SRB as follows
\begin{equation}
    \Bar{r}_{m}(s_t, \Bar{a}_{t,m}) = \frac{1}{K}\sum_{n\in\mathcal{N}}\sum_{f\in\mathcal{F}_{n,m}}\sum_{k\in\mathcal{K}_n}R_{k,n,f}(s_t, \Bar{a}_{t}).
\end{equation}

\subsubsection{Action space} For subchannel assignment, let define the action space $\mathcal{A}_{n,m} = (\mathcal{K}_n \cup \{0\})^{|\mathcal{F}_{n,m}|}$, where $\mathcal{K}_n$ is the set of users associated with BS $n$ and the additional action~0 represents the `{unactivated}' action (of a subchannel). It is worth noting that the inclusion of the last action into the action space leads to significant and important effects in \emph{interference mitigation}. For example, the each agent can arrange with other agents to deactivate certain subchannels to mitigate inter-cell interference. Moreover, by deactivating these likely weak subchannels, the transmit power is concentrated on the stronger activated subchannels, resulting in higher efficiency in terms of both power and spectrum.

For power allocation, we propose an indirect approach via subchannel assignment. First, each BS equally allocates the total power to its SRBs. Within each SRB, the power is uniformly allocated to \emph{only active subchannels}. As a result, the transmit power on each subchannel can take a value from $\{p_{\max}/M, p_{\max}/2M, \ldots, p_{\max}/|\mathcal{F}_{n,m}|M\}$, corresponding to the cases where one, two, \ldots, or all subchannels are activated. 

\begin{claim} 
    The designs of reward functions and actions space presented above indeed satisfy both Assumptions \ref{assumption:reward_invariant} and \ref{assumption:reward_decomposition}.
\end{claim}
To verify Claim~1, we first see that there is no interference between different SRBs of the same BS. Second, under conditions in~\eqref{eq:decomposition}, there is no interference between different SRBs controlled by agents from different agent groups. Third, the proposed power allocation approach guaranties that the action of each agent group does not affect the underlying reward function of other agent groups.

\subsubsection{Observation} The local observation $\Bar{z}_{t,n,m}$ of agent $A_{n,m}$ in time step $t$ includes: $i$) channel gains of subchannels in $\mathcal{F}_{n,m}$ to the associated users in $\mathcal{K}_n$, $ii$) pathloss of all BSs to all users, and $iii$) the approximation of accumulated rate achieved by the preceding $m-1$ agents of BS $n$, i.e., $\{\Psi_{n,m,k}:k\in \mathcal{K}_n\}$ where $\Psi_{n,m,k} = \sum_{m'=1}^{m-1}\sum_{f\in\mathcal{F}_{n,m'}} R_{k,n,f}(s_t, a_t).$
With the last information, agent $A_{n,m}$ can approximately assess the achievements of preceding agents and adjust its actions to balance the main objective with constraint satisfaction.

\subsubsection{State space} the state $s_t$ includes channel gains and pathloss of the whole system in the time step $t$. Note that $s_t$ is only partially observed each agent.

\subsubsection{Cost function} the cost function is designed to guide the agents toward feasible regions of the policy space. The cost function of user $k$ associated with BS $n$ for taking the joint action $\Bar{a}_t$ in the state $s_t$ is defined as 
\begin{equation}
    c_k(s_t, \Bar{a}_t) = \max\big(\eta_k - {\sum}_{f\in\mathcal{F}_n}R_{k,n,f}(s_t, \Bar{a}_t), 0\big).\label{eq:cost_function}
\end{equation}
An agent receives a positive cost value if the QoS constraint of user $k$ is violated, and a cost value of zero otherwise. Negative cost values are truncated to reduce their effects in constraint handling, as we will present shortly.

\subsection{Constraint Handling}
Let $d_{\Bar{\pi},\mu}$ be the discounted state-occupancy measure under the joint policy $\Bar{\pi}$ and initial state distribution $\mu$. We have the following useful result:
\begin{proposition}
    \label{proposition:discounted_to_avg}
    For any function $f:\mathcal{S}\times\mathcal{A}\rightarrow \mathbb{R}$, we have
    \begin{equation}
        \mathbb{E}_{\Bar{\pi},\mu}\Big[\sum_{t=0}^{\infty}\gamma^t f(s_t, a_t)\Big] = \frac{1}{1-\gamma}\mathbb{E}_{s\sim d_{\Bar{\pi}, \mu}, a\sim\Bar{\pi}}\left[f(s, a)\right].
        \label{eq:discounted_to_avg}
    \end{equation}
\end{proposition}
With the cost function \eqref{eq:cost_function}, we can safely relax \eqref{prob:cmdp_decomposed} by replacing constraint \eqref{ctr:qos_cmdp_decomp} by the following new constraint:
\begin{equation}
    \mathbb{E}_{s\sim d_{\Bar{\pi}, \mu}, \Bar{a}_m\sim\Bar{\pi}_m}\left[c_{k}(s, \Bar{a})|\Bar{\pi}_{1:m-1}\right] \leq 0, \forall k\in \mathcal{K},\label{ctr:avg_cost}
\end{equation}
where $\Bar{a}_m$ is a joint action of agent group $m$.
Here, we have replaced the stepwise constraint with a long-term average constraint and set $d_k$ to zero. Since negative cost values are all truncated, a policy that satisfies $\eqref{ctr:avg_cost}$ is guaranteed to satisfy \eqref{ctr:qos_cmdp_decomp}. By proposition \ref{proposition:discounted_to_avg}, we can rewrite \eqref{ctr:avg_cost} as
\begin{equation}
    (1-\gamma)\mathbb{E}_{\Bar{\pi}_m,\mu}\Big[\sum_{t=0}^{\infty}\gamma^t c_k(s_t, \Bar{a}_t) |\Bar{\pi}_{1:m-1}\Big] \leq 0, \forall k\in \mathcal{K}.\label{ctr:discounted_cost}
\end{equation}
Assume that $\Bar{\pi}_m$ is parameterized by $\theta_m$, we use the technique of Lagrangian relaxation to migrate constraint \eqref{ctr:discounted_cost} into the objective function of \eqref{prob:cmdp_decomposed}:
\begin{align}
    &\underset{\boldsymbol{\lambda}}{\min}~\underset{\theta_m}{\max}\quad \mathcal{L}(\boldsymbol{\lambda}, \theta_m) = \mathbb{E}_{\Bar{\pi}_m,\mu}\Big[\sum_{t=0}^{\infty}\gamma^t\Bar{r}_m(s_t, \Bar{a}_{t,m})|\Bar{\pi}_{1:m-1}\Big] \nonumber\\
    &\quad-\sum_{k\in\mathcal{K}}\lambda_k(1-\gamma)\mathbb{E}_{\Bar{\pi}_m,\mu}\Big[\sum_{t=0}^{\infty}\gamma^t c_k(s_t, \Bar{a}_t) |\Bar{\pi}_{1:m-1}\Big].\label{prob:cmdp_relax_unconstrained_long}
\end{align}
where $\boldsymbol{\lambda}$ are non-negative Lagrange multipliers. By denoting $\hat{r}_m(s_t, \Bar{a}_{t,m}; \boldsymbol{\lambda}) = \Bar{r}_m(s_t, \Bar{a}_{t,m}) - \sum_{k\in\mathcal{K}}\lambda_k(1-\gamma)c_k(s_t, \Bar{a}_t)$, we obtain an equivalent problem as follows:
\begin{align}
    \underset{\boldsymbol{\lambda}}{\min}~\underset{\theta_m}{\max}~\mathbb{E}_{\Bar{\pi}_m,\mu}\left[{\sum}_{t=0}^{\infty}\gamma^t\hat{r}_m(s_t, \Bar{a}_{t,m})|\Bar{\pi}_{1:m-1}\right]\label{prob:cmdp_relax_unconstrained}.
\end{align}
Observe that optimizing $\theta_m$ for a fixed $\boldsymbol{\lambda}$ is exactly the same as solving an unconstrained reinforcement learning problem, and that, as $\boldsymbol{\lambda}$ increases, the solution of \eqref{prob:cmdp_relax_unconstrained} converges to a feasible solution of \eqref{prob:cmdp_decomposed}. These observations suggest a learning procedure in which we can adopt an unconstrained learning algorithm to solve \eqref{prob:cmdp_relax_unconstrained} for a given $\boldsymbol{\lambda}$, while simultaneously gradually increasing $\boldsymbol{\lambda}$ at a slower timescale until all constraints are satisfied. This results in an iterative procedure, where $\theta_m$ can be updated as in any policy gradient algorithms and $\boldsymbol{\lambda}$ is updated using sub-gradients:
\begin{align}
    \boldsymbol{\lambda}' = \boldsymbol{\lambda} - \alpha\nabla_{\boldsymbol{\lambda}}\mathcal{L}(\boldsymbol{\lambda}, \theta_m)\label{eq:update_lagr_mult}
\end{align}
where $\alpha$ is the learning rate of $\boldsymbol{\lambda}$ and  $\nabla_{\boldsymbol{\lambda}}\mathcal{L}(\boldsymbol{\lambda}, \theta_m)$ can be derived from \eqref{prob:cmdp_relax_unconstrained_long} as follows
\begin{align}
    \nabla_{\boldsymbol{\lambda}_k}\mathcal{L}(\boldsymbol{\lambda}, \theta_m) =-\mathbb{E}_{s\sim d_{\Bar{\pi}, \mu}, \Bar{a}_m\sim \Bar{\pi}_m}\left[c_{k}(s, \Bar{a})|\Bar{\pi}_{1:m-1}\right],
\end{align}
where we have applied proposition \ref{proposition:discounted_to_avg} to convert the gradient into the expectation over the state-action occupancy measure for convenience of estimation during training.

\subsection{Enhance Flexibility via Dynamic Environment Training}
The $M$ agents of each BS are trained to jointly serve $K_{\max}$ users, which is usually limited by the total subchannels. To improve the robustness of the proposed policy, during the training, we incorporate the system dynamics to allow an arbitrary number of active users ($\leq K_{max}$). In particular, while the maximum number of users $K_{\max}$ is fixed, the number of active users and their locations are randomly generated in each training episode following a predefined random process. 
Moreover, to provide the agents with a better understanding of the system, metadata about user presence in the system is included in the agents' observations and the environment state. Specifically, let $\psi_{t,n} = \{\psi_{t,n,k} : k \in \mathcal{K}\}$ be the metadata observed by the agents of BS $n$ at time step $t$. Here, $\psi_{n,k,t} = 1$ if an active user $k$ occupies a subchannel at time step $t$, and $\psi_{n,k,t} = 0$ otherwise. By incorporating $\psi_{t,n}$ into the observations, the active status of users is treated as an input to the policies, rather than as an underlying assumption of the environment, thereby enhancing the flexibility of the learned policies.

\section{Simulation Results}
\label{sec:experiments}
The proposed algorithm is trained and validated in a system consisting of 2 TBSs and one UAV (NTBS) hovering at 200m height, covering an area of 1000m$\times$1000m. 
All BSs share a common frequency band of 20 MHz. Each BS has  a maximum transmit power of 46dBm. Different numerologies are tested, resulting in a total subchannels of $\{80, 160, 320\}$. The number of active users at each BS follows the Poisson distribution with a mean of $\lambda = 0.5K_{\max}$. The minimum data rate of each user is set to 2 Mbps. We use the standard Rayleigh fading channel model for TBS-user, and probabilistic model for NTBS-user links, with the Winner II pathloss model. 

We adopt the multi-agent PPO (MAPPO) algorithm \cite{yu2022surprising} as the computational workhorse of our proposed solution. Each BS are served by multiple agents, each of which manages a SRB of 40 subchannels. Each training episode includes $100$ steps, corresponding to 100 frames, or 1 second in 5G systems. 
We compare our solution with the two most popular approaches: MAPPO and IPPO \cite{naderializadeh2021resource, nasir2019multi}, which  apply the MAPPO and Independent PPO algorithms, respectively, to solve \eqref{prob:cmdp} directly without decomposition. Details of these algorithms can be found in \cite{yu2022surprising}. The neural architectures and parameters of all three solutions are identical across all experiments. We use a learning rate of $5e^{-5}$, a discount factor equals to $0.9$, a batch size of 64, a replay buffer size of 105 samples, and a target network update rate of 0.01.

\begin{figure}
     \centering
     \begin{subfigure}{0.32\columnwidth}
         \centering
         \includegraphics[width=\textwidth]{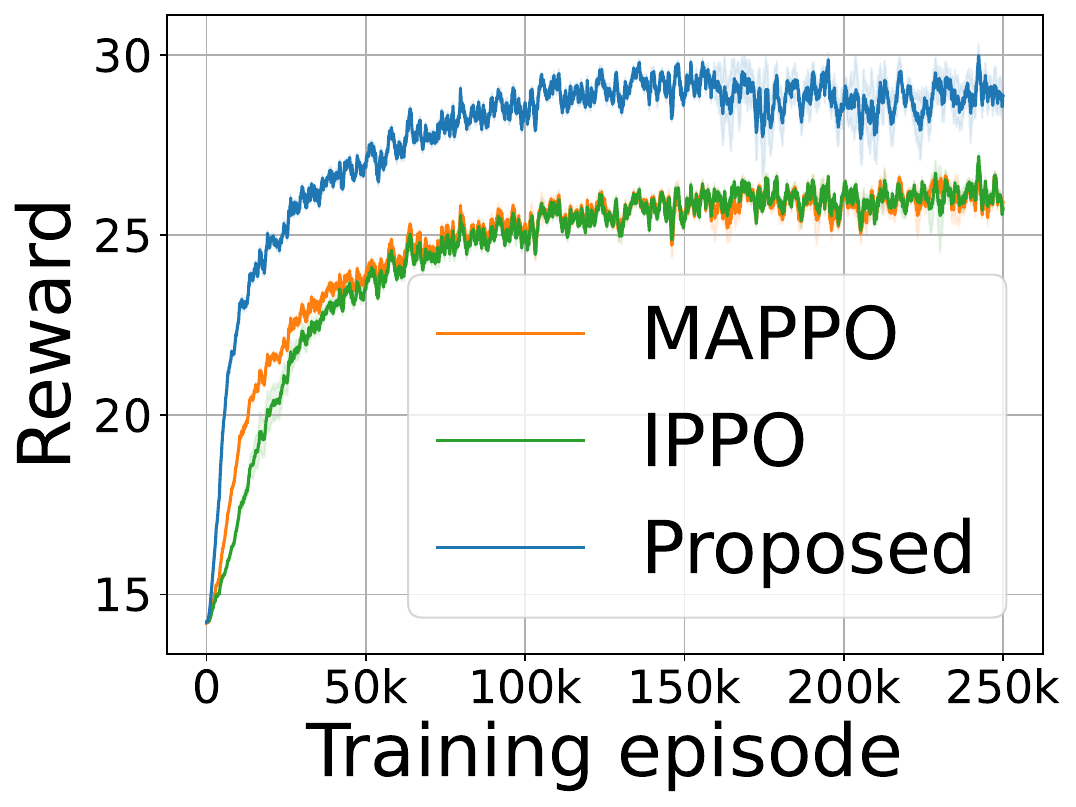}
         \caption{30 UEs, 240 channels}
     \end{subfigure}
     \hfill
     \begin{subfigure}{0.31\columnwidth}
         \centering
         \includegraphics[width=\textwidth]{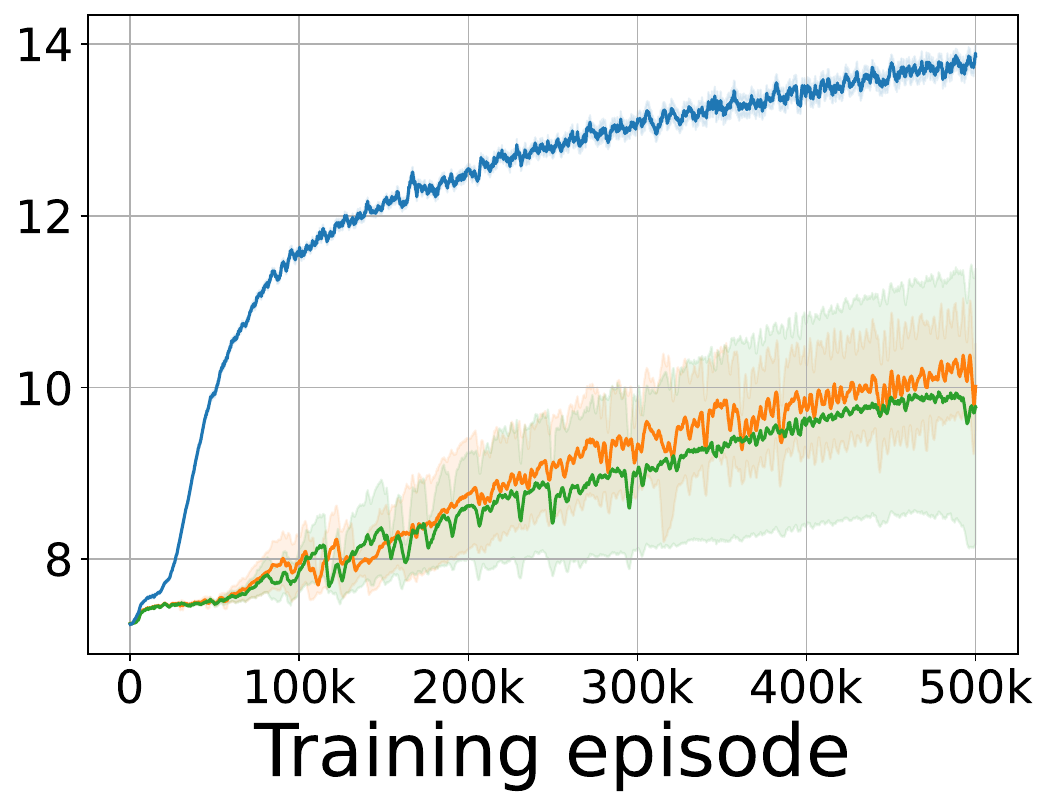}
         \caption{60 UEs, 480 channels}
     \end{subfigure}
     \hfill
     \begin{subfigure}{0.3\columnwidth}
         \centering
         \includegraphics[width=\textwidth]{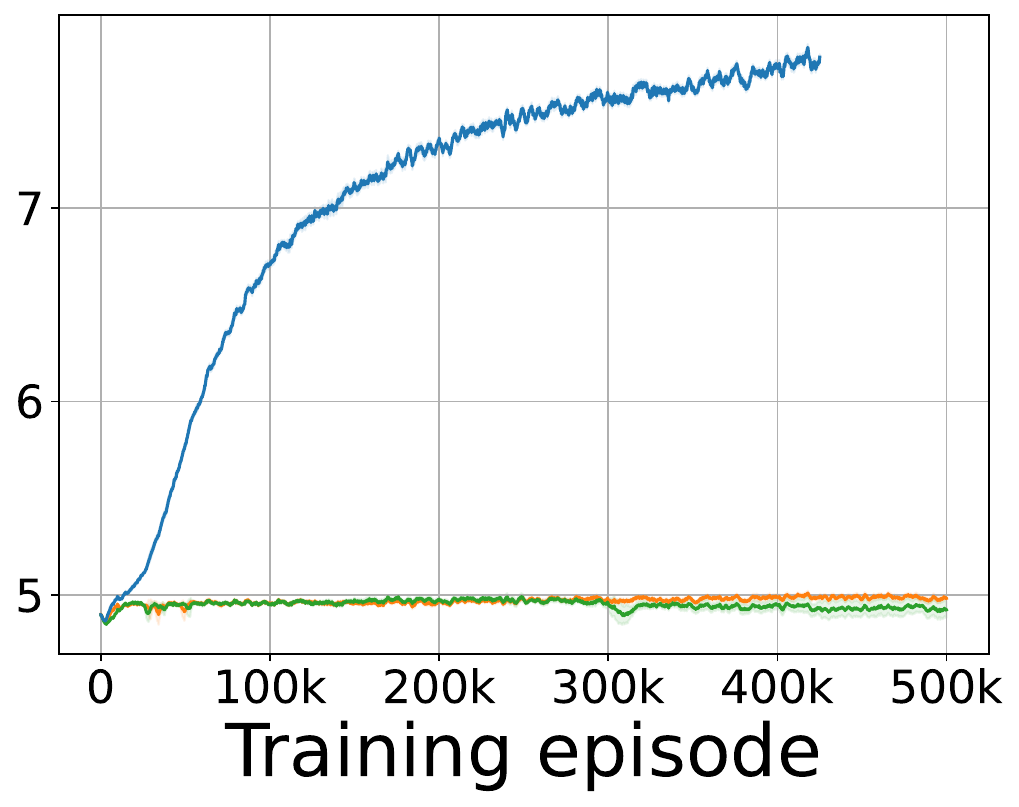}
         \caption{90 UEs, 960 channels}
     \end{subfigure}
     \hfill
     \begin{subfigure}{0.32\columnwidth}
         \centering
         \includegraphics[width=\textwidth]{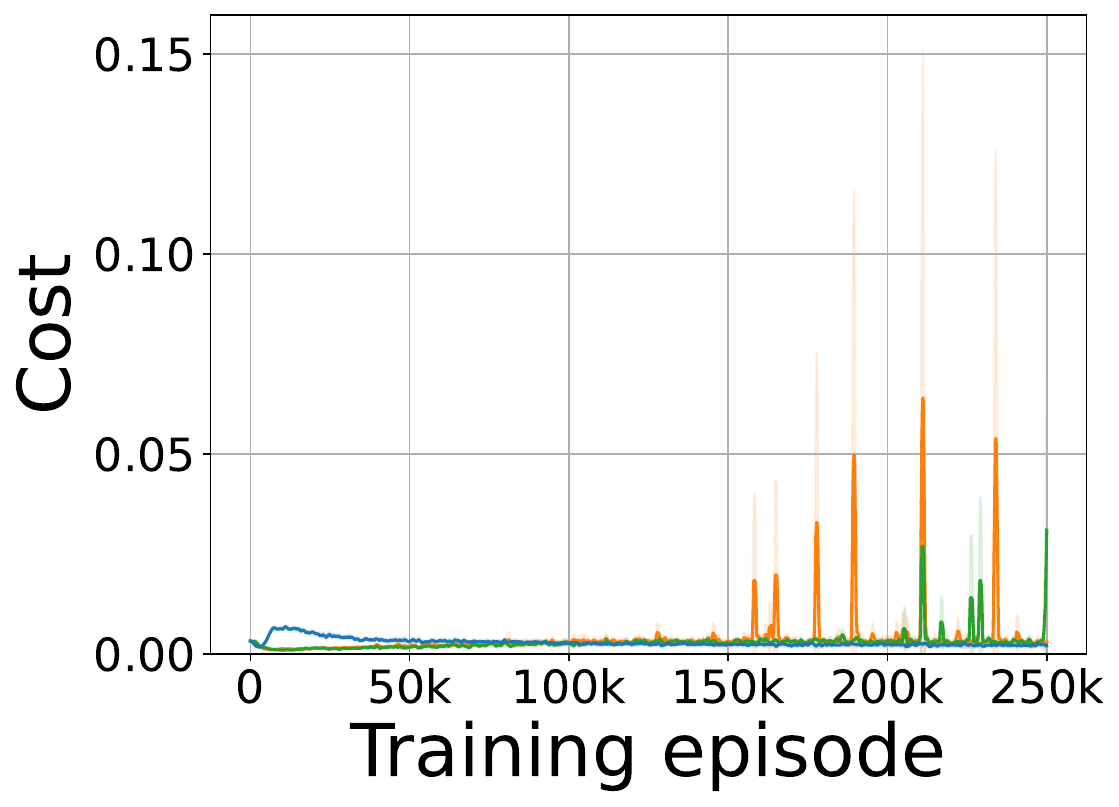}
         \caption{30 UEs, 240 channels}
     \end{subfigure}
     \hfill
     \begin{subfigure}{0.32\columnwidth}
         \centering
         \includegraphics[width=\textwidth]{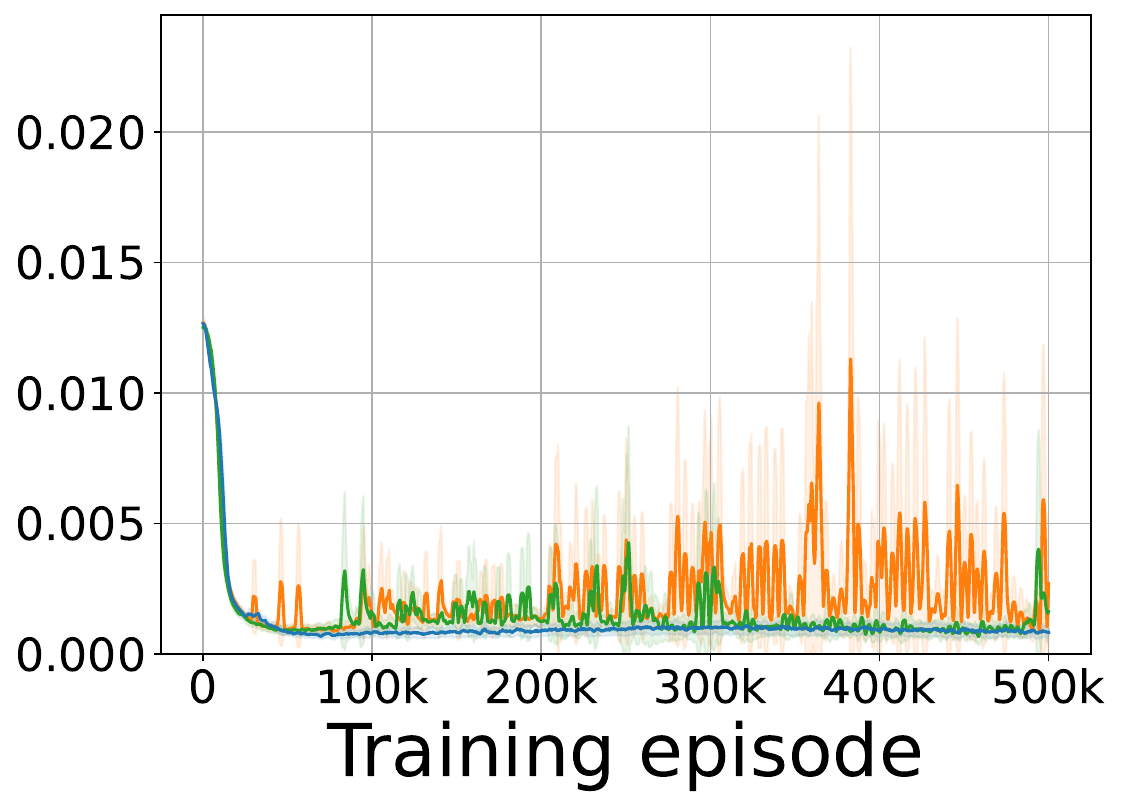}
         \caption{60 UEs, 480 channels}
     \end{subfigure}
     \hfill
     \begin{subfigure}{0.3\columnwidth}
         \centering
         \includegraphics[width=\textwidth]{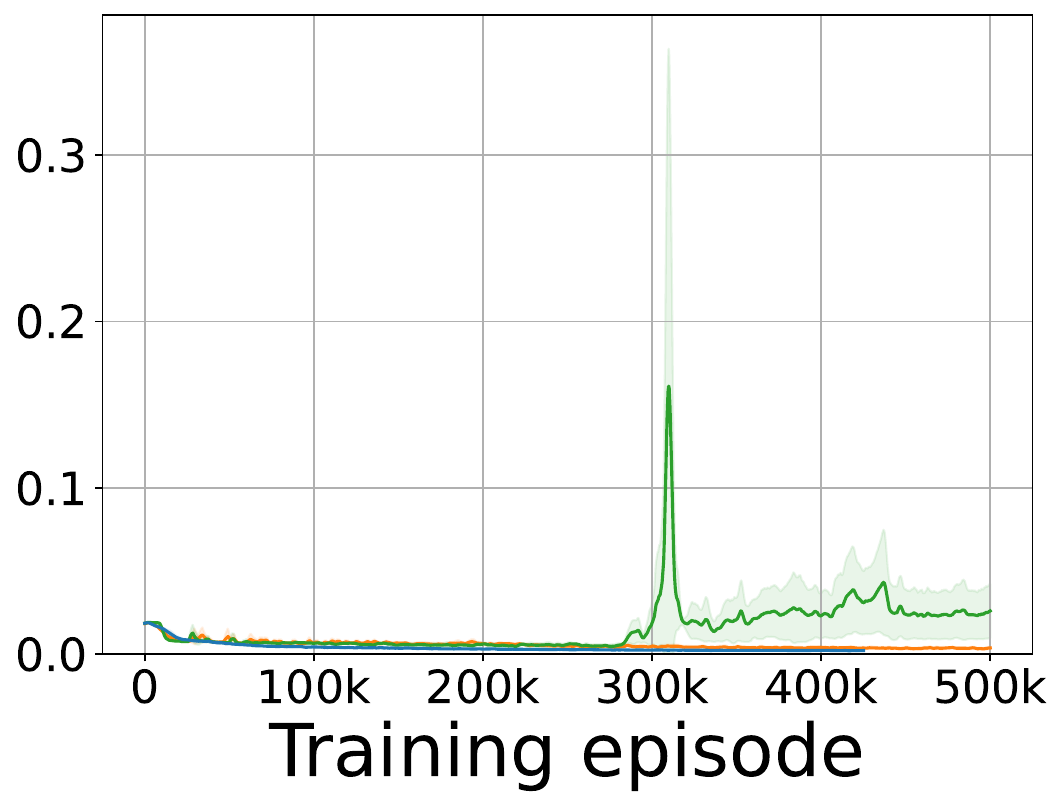}
         \caption{90 UEs, 960 channels}
     \end{subfigure}
     \caption{The scalability of all solutions demonstrated by the learning performance on different number of users and subchannels.}
     \label{fig:learning_performance}
\end{figure}

\textbf{Scalability}: Fig. \ref{fig:learning_performance} demonstrates the learning performance of the algorithms under different subcarrier spacings and $K_{\max}$. Specifically, the agents of each BS are trained to manage 80, 160, and 320 subchannels to serve a maximum of 10, 20, and 30 users, respectively. This results in total numbers of 30, 60, and 90 users, and 240, 480, and 960 subchannels in the system. Fig. \ref{fig:learning_performance} shows that the proposed solution achieves the highest reward and the smallest cost of violating the constraints in all settings. Moreover, while MAPPO and IPPO fail to learn in the latter two settings, the proposed solution continues to improve the reward while maintaining a minimum constraint violation. The results prove the robustness and scalability of the proposed algorithm against the problem size.

\begin{figure}
     \centering
     \begin{subfigure}{0.32\columnwidth}
         \centering
         \includegraphics[width=\textwidth]{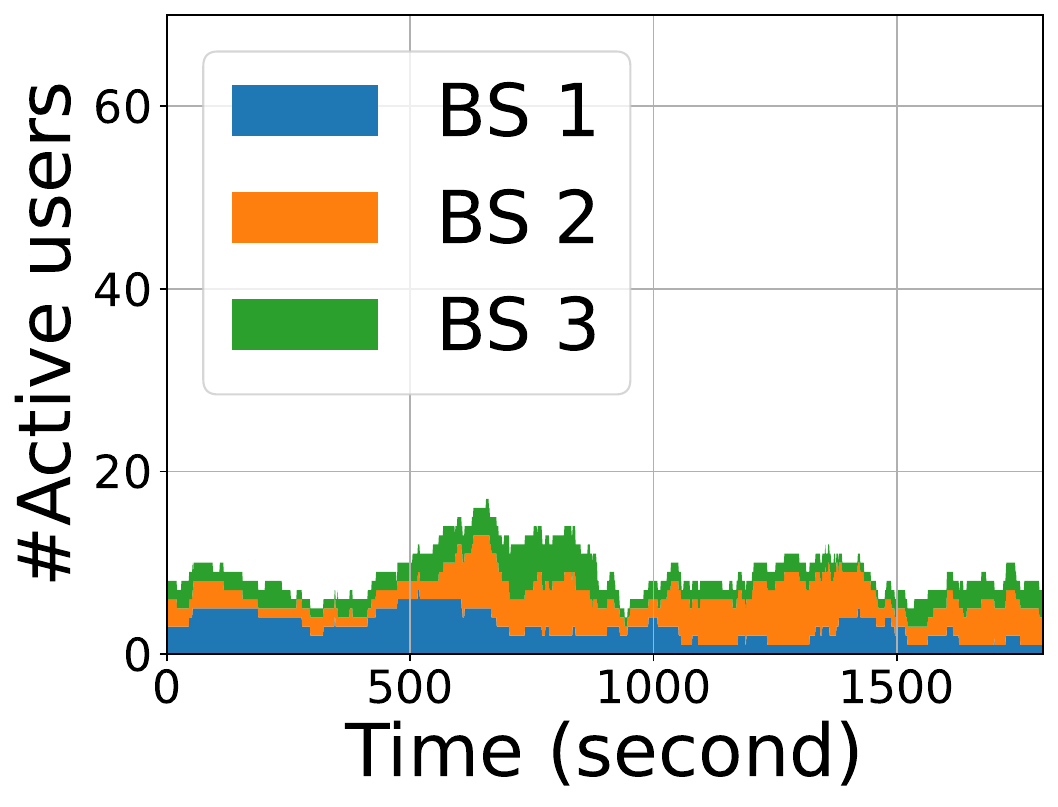}
         \caption{1 UE/min./BS}
     \end{subfigure}
     \hfill
     \begin{subfigure}{0.31\columnwidth}
         \centering
         \includegraphics[width=\textwidth]{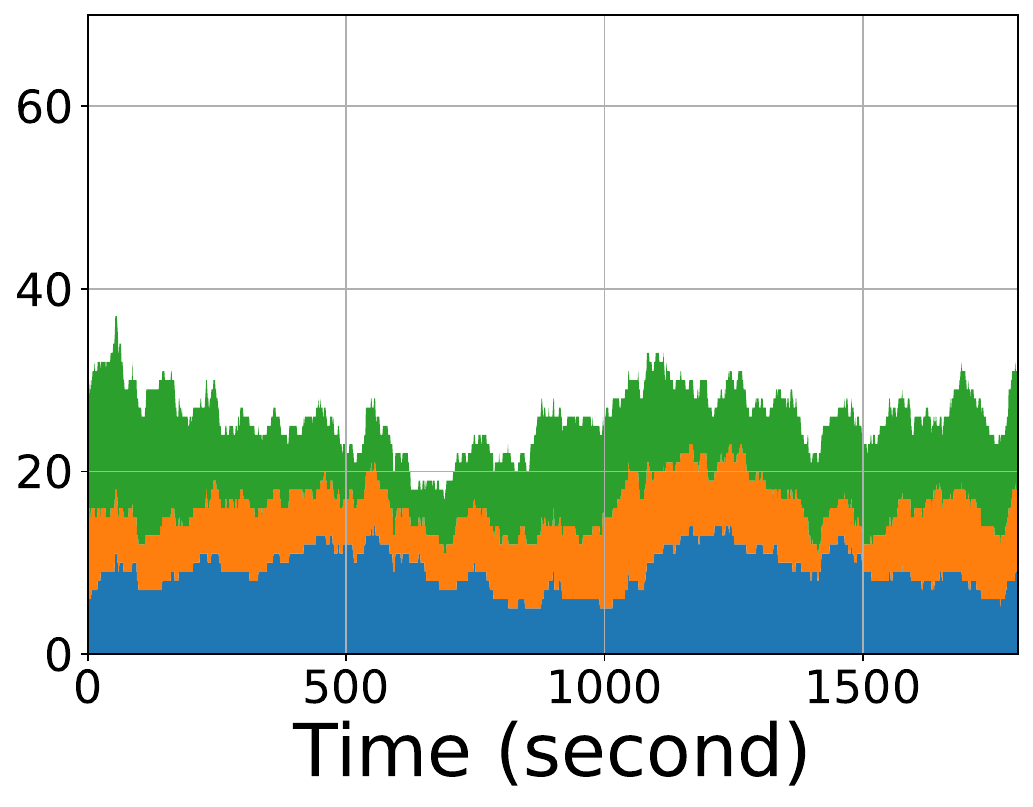}
         \caption{3 UEs/min./BS}
     \end{subfigure}
     \hfill
     \begin{subfigure}{0.31\columnwidth}
         \centering
         \includegraphics[width=\textwidth]{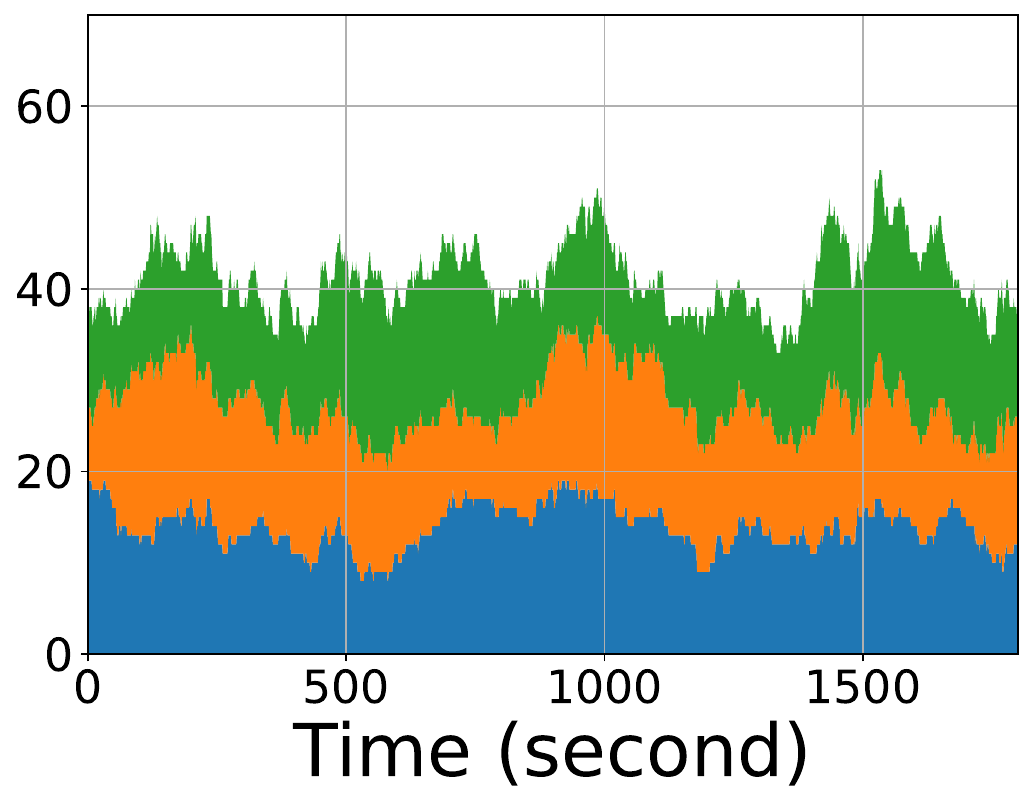}
         \caption{5 UEs/min./BS}
     \end{subfigure}
     \hfill
     \begin{subfigure}{0.31\columnwidth}
         \centering
         \includegraphics[width=\textwidth]{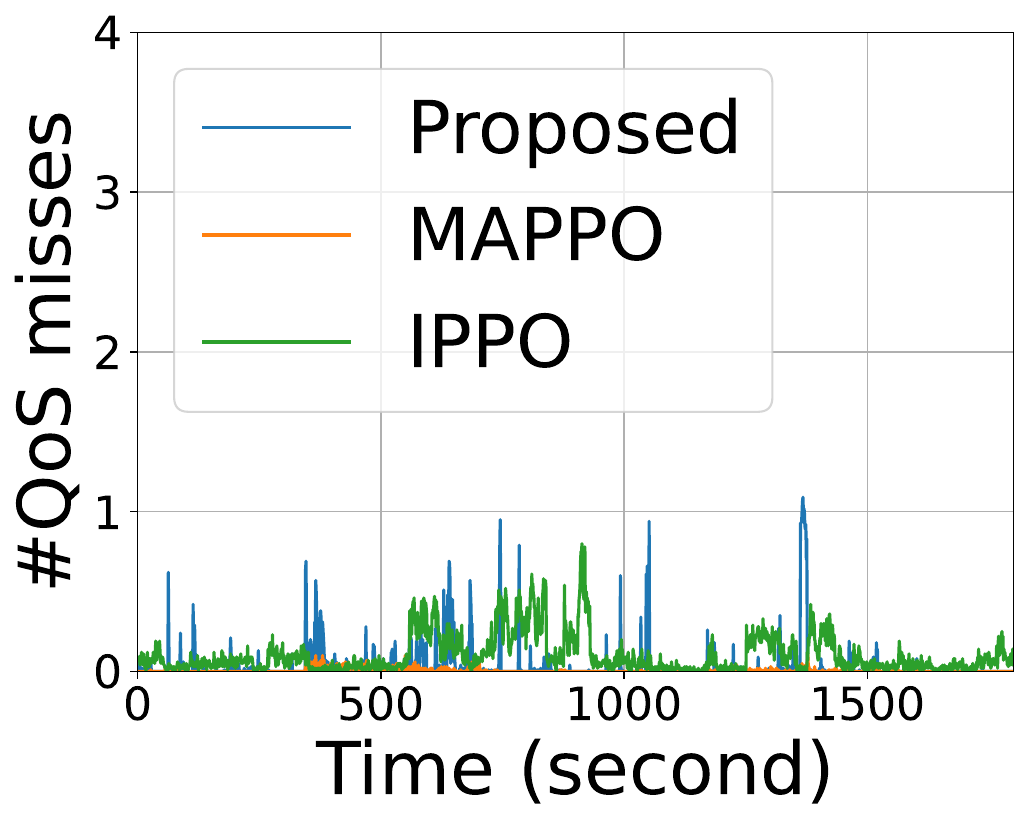}
         \caption{1 UE/min./BS}
     \end{subfigure}
     \hfill
     \begin{subfigure}{0.31\columnwidth}
         \centering
         \includegraphics[width=\textwidth]{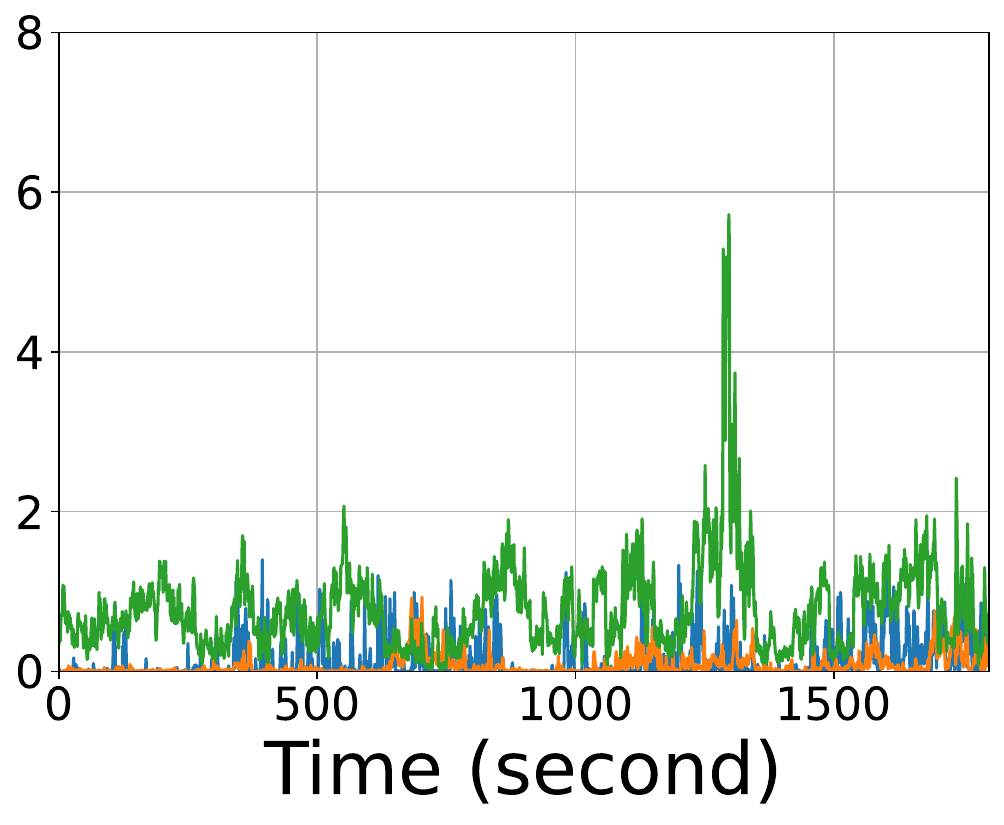}
         \caption{3 UEs/min./BS}
     \end{subfigure}
     \hfill
     \begin{subfigure}{0.31\columnwidth}
         \centering
         \includegraphics[width=\textwidth]{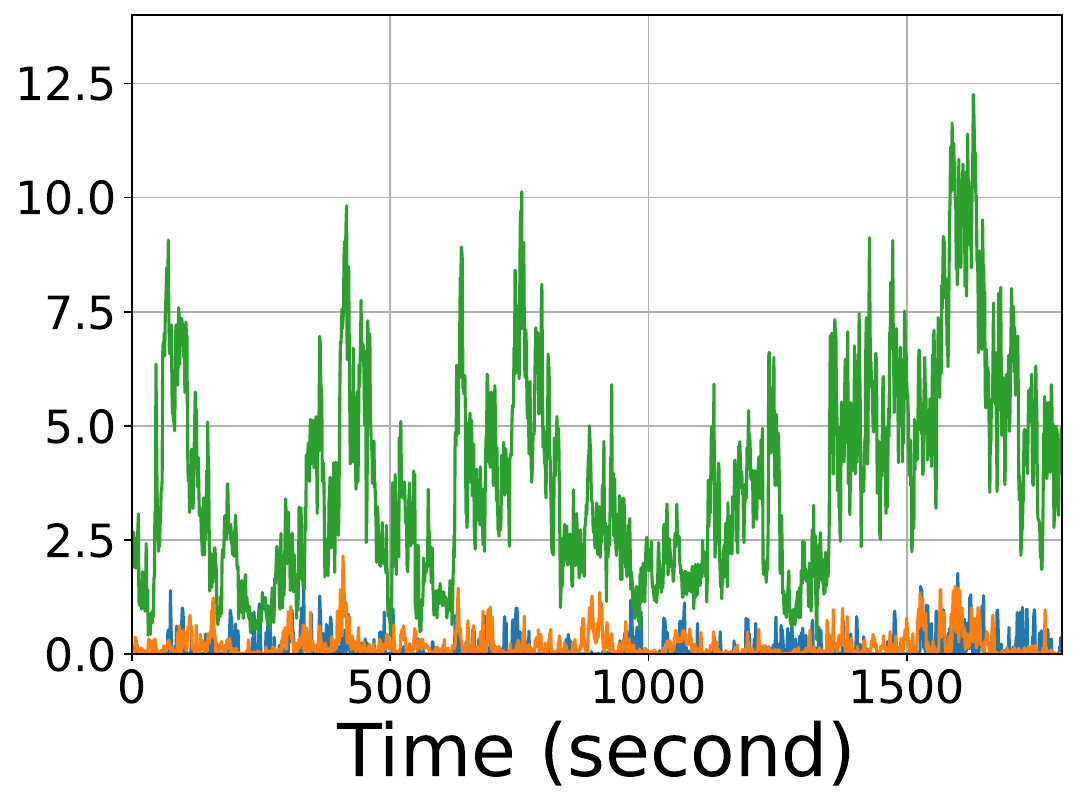}
         \caption{5 UEs/min./BS}
     \end{subfigure}
     \hfill
     \begin{subfigure}{0.31\columnwidth}
         \centering
         \includegraphics[width=\textwidth]{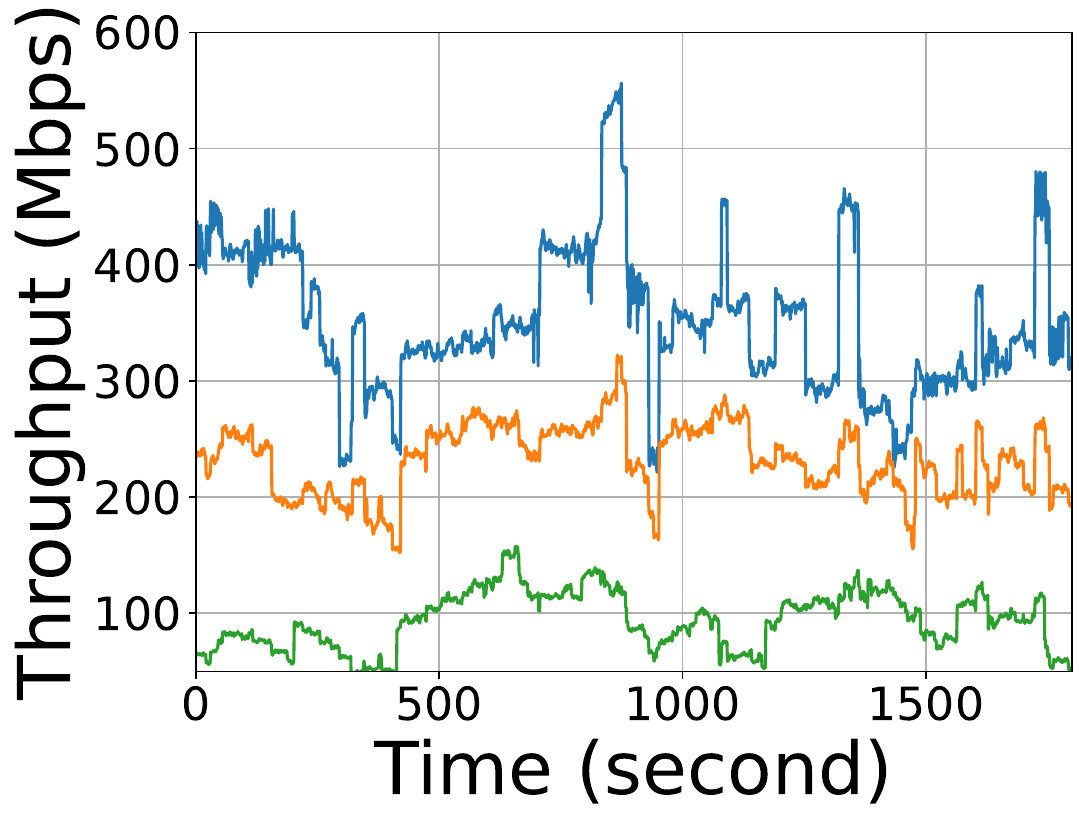}
         \caption{1 UE/min./BS}
     \end{subfigure}
     \hfill
     \begin{subfigure}{0.31\columnwidth}
         \centering
         \includegraphics[width=\textwidth]{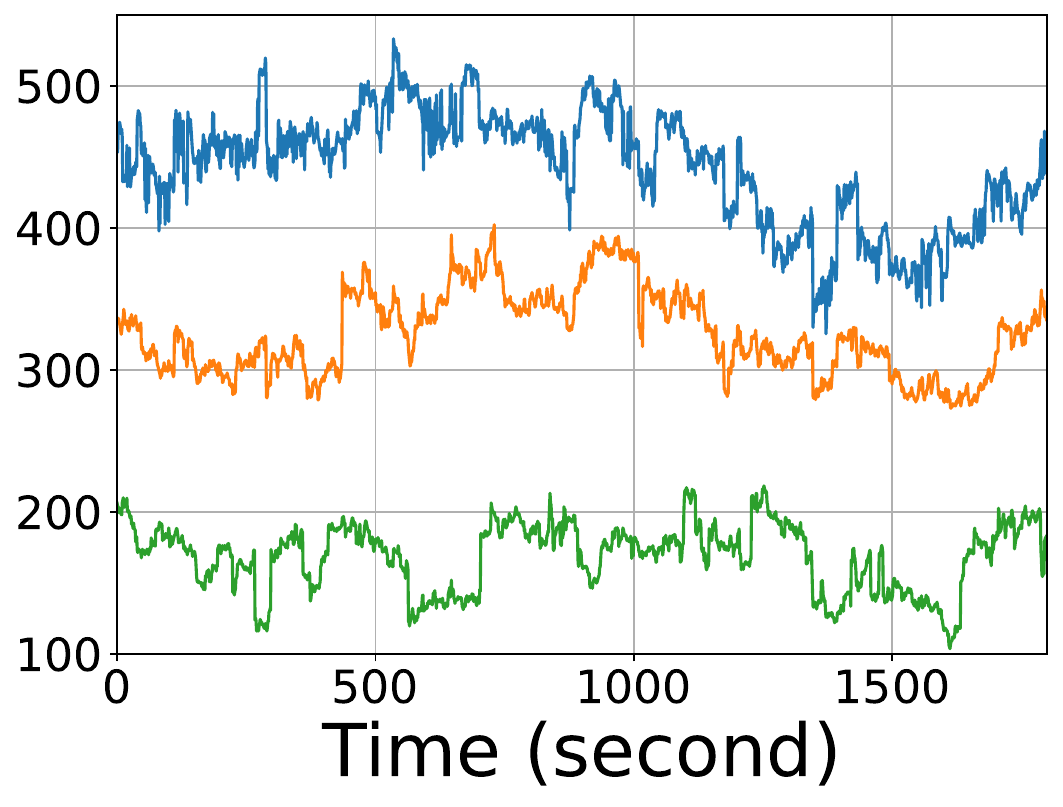}
         \caption{3 UEs/min./BS}
     \end{subfigure}
     \hfill
     \begin{subfigure}{0.31\columnwidth}
         \centering
         \includegraphics[width=\textwidth]{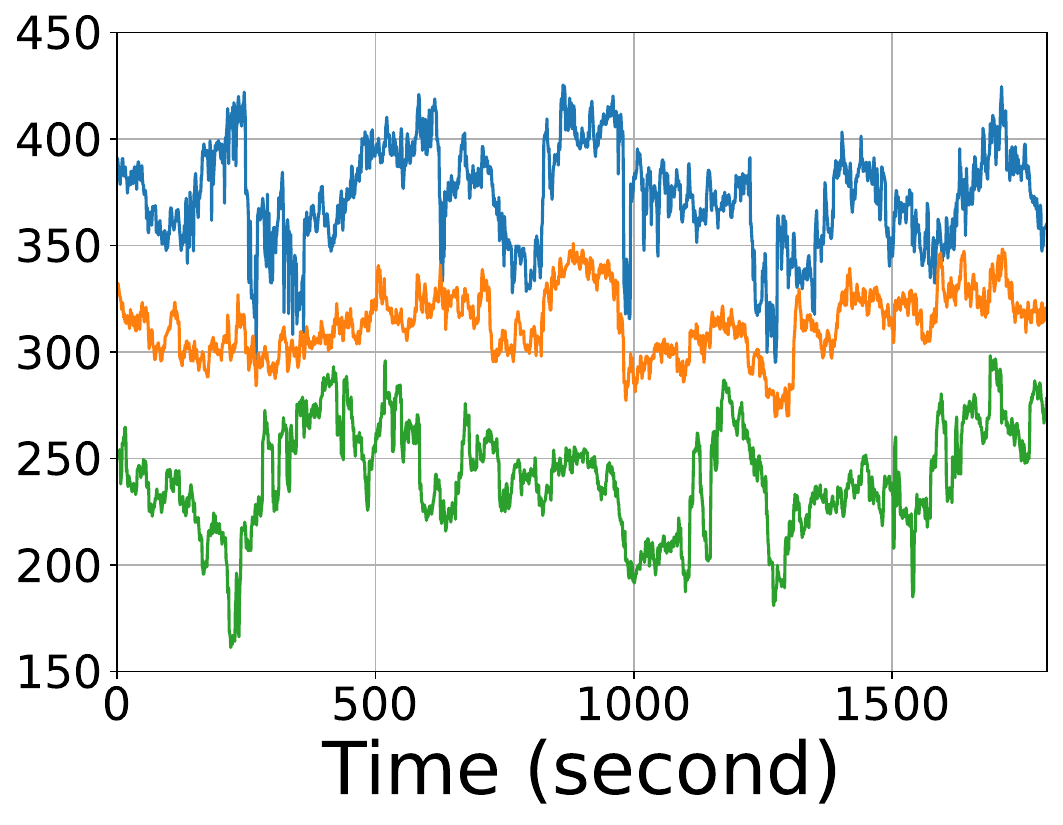}
         \caption{5 UEs/min./BS}
     \end{subfigure}
     
     \caption{The flexibility and robustness of learned policies demonstrated by testing results on a real-world scenario.}
     \label{fig:testing}
\end{figure}

\textbf{Flexibility}: Fig. \ref{fig:testing} depicts the testing results of the policies under the second setting (160 subchannels to serve a maximum of 20 users at each BS) for a duration of 30 minutes. New users arrive at each BS every at a rate of 1, 3, or 5 per minute, and their active duration follows an exponential distribution with a mean of 3 minutes. Users mobility follows random walks with a speed of 1 m/s. The actual number of active users are plotted in Fig. \ref{fig:testing}(a) - (c).  The advantages of the proposed learning algorithm are clearly shown via plot the number of active users, average number of QoS misses, and total system throughput, respectively, over time. As shown in this figure, the proposed solution achieves the highest system throughput among all solutions while maintaining low QoS violations, even as user demand increases.

\section{Conclusion}
\label{sec:conclusion}
The decentralization requirement of TN-NTN systems demands efficient multi-agent learning approaches for complex resource allocation. This study provides insights into addressing scalability and flexibility challenges in real-world scenarios. By exploiting the system's unique characteristics, we propose a decomposition approach with minimal performance loss and a dynamic training environment to enhance policy flexibility. Experiments on very large-scale settings of joint TN-NTN resource allocation confirm the superiority of the proposed solution in terms of scalability, flexibility, and robustness in dynamic environments.


\end{document}